\documentclass[10pt,journal,twoside]{IEEEtran}
\usepackage{multirow}
\usepackage{amssymb}
\usepackage[dvips]{graphicx}
\usepackage{amsmath}
\usepackage{amsthm}
\usepackage{latexsym,bm}
\usepackage{color}
\usepackage{subfigure}
\usepackage{longtable}
\usepackage{accents}
\usepackage{cite}
\usepackage{enumerate}
\usepackage{arydshln}
\usepackage{slashbox}
\usepackage{algorithmic}
\usepackage{float}
\makeatletter
\def\widebar{\accentset{{\cc@style\underline{\mskip10mu}}}}
\def\Widebar{\accentset{{\cc@style\underline{\mskip8mu}}}}
\makeatother

\theoremstyle{plain}

\theoremstyle{definition}
\theoremstyle{definition}

\begin{document}

\title{ Physical-Layer Network Coding: An Efficient Technique for Wireless Communications
\thanks{ Pingping Chen is with the Fuzhou University  and the National Mobile Communications Research Laboratory, Southeast University; Zhaopeng Xie and Zhifeng Chen are with the Fuzhou University; Yi~Fang (Corresponding author) is with the Guangdong University of Technology and the Fuzhou University; Shahid Mumtaz is with the Instituto de Telecomunica\c{c}\~oes, Portugal; Joel J. P. C. Rodrigues is with the National Institute of Telecommunications (Inatel), Santa Rita do Sapuca\'i, MG, Brazil, Instituto de Telecomunica\c{c}\~oes, Portugal, and Federal University of Piau\'i (UFPI), Teresina - PI, Brazil.
}
}


\author{\normalsize  Pingping Chen, Zhaopeng Xie, Yi Fang,  Zhifeng Chen,   Shahid Mumtaz, and Joel J. P. C. Rodrigues
}


\markboth{IEEE Network Magazine, Accepted for Publication (Paper Number: NETWORK-19-00289), July 2019}%
{Chen \MakeLowercase{\textit{et al.}}: Physical-Layer Network Coding: An Efficient Technique for Wireless Communications}

\maketitle
\begin{abstract}

As a subfield of network coding, physical-layer network coding (PNC)  can  effectively enhance the throughput of
  wireless networks by mapping superimposed signals at receiver  to other forms of user messages.
Over the past twenty years, PNC has received significant research attention  and has been widely studied in various   communication scenarios, e.g., two-way relay communications (TWRC),  nonorthogonal
multiple access (NOMA) in 5G networks, random access networks, etc. Later on, channel-coded PNC and related  communication techniques were investigated to ensure network reliability, such as the design of channel code, low-complexity decoding, and cross-layer design.  In this article, we briefly review  the variants of channel-coded PNC-aided wireless communications with the aim of
inspiring future research activities in this area. We also put forth  open research problems along with a few selected research directions under PNC-aided frameworks.

%
%
%
%
\end{abstract}


\section{Introduction}
In the fifth generation (5G) wireless communication networks,
non-orthogonal multiple access (NOMA) has been attracting
significant research efforts  for the design of radio
access techniques \cite{5G}. However, NOMA suffers from degraded spectral efficiency due to  signal interferences in simultaneous user transmissions. To overcome this problem, many coding and signal
processing techniques were proposed to mitigate and utilize the multiuser interference.
The concept of physical-layer network coding (PNC), inspired by traditional network coding, was proposed and demonstrated its advantages over the conventional communications from both
 information-theoretic and practical perspectives \cite{PNC}. In particular, PNC can significantly improve the network throughput by exploiting the characteristic of user interferences. {Moreover, those design schemes based on PNC are becoming  competitive solutions for NOMA in 5G networks \cite{NOMA}.}
In  PNC-aided networks, the receiver is dedicated to decoding linearly weighted combinations of   user messages, referred as network-coded (NC) message, from the received signals. {A simple PNC operated network is  two-way relay channel (TWRC), in which two user nodes
desire to communication with each other via a relay.  A PNC-aided TWRC  has two phases. The first phase is multiple access phase and the second one is broadcast phase.  As illustrated in Fig. 1, in the first phase, user 1 sends message $S_1$ and user 2 sends message $S_2$ to the relay simultaneously. Given the superimposed message  from two users, the relay attempts to decode a linear combination of $S_1$ and $S_2$, $S_1\oplus S_2$,  Then, in the second phase,  the relay broadcasts $S_1\oplus S_2$  to the two users.} By doing so, PNC was shown to  outperform the conventional transmission scheme in practical scenarios in terms of sum-rate  and decoding performance \cite{LIEW,NAZ}.

\begin{figure}[!t]
\centerline{
\includegraphics[width=3.3in,height=1.8in]{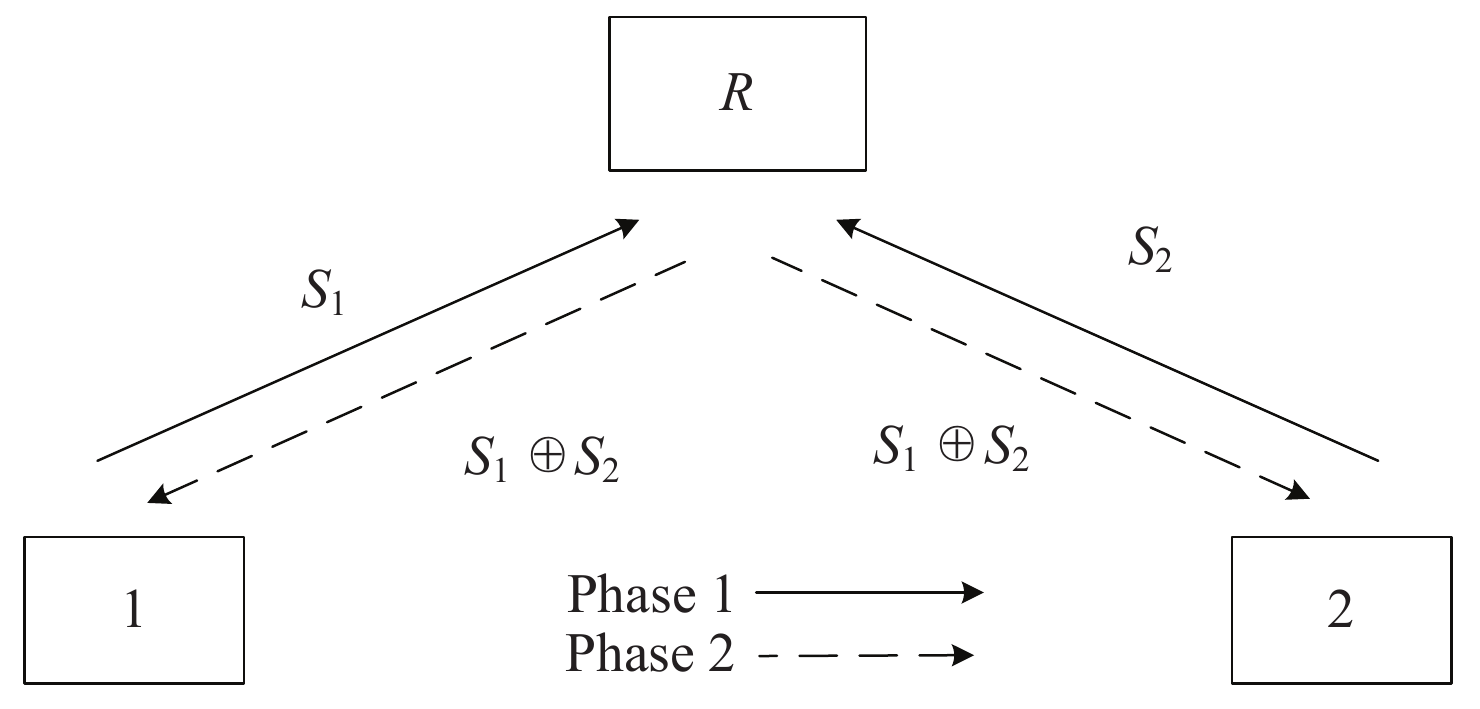}}
  \caption{Physical-layer network coding.}
  \label{fig:Fig1}
\end{figure}
%
%
 {As an emerging technique,  PNC brings up many issues  in
a variety of  wireless communications scenarios. Interesting
 issues include how to  characterize the decoding behavior
of PNC  in power-imbalance TWRC \cite{LONG1}, and how to attain full-rate
and full-diversity in PNC-aided  multiple-input multiple-output (MIMO) scenarios  \cite{MIMO}.
 An information-theoretic issue is how  to design powerful channel coding to approach the PNC cut-set bound on the information capacity.
The lattice-coded PNC has been shown to achieve a small gap less than 1/2 from the cut-set bound in power-imbalanced Gaussian channels \cite{LATTICE}.
 However, high
encoding/decoding complexities  hinder the practical implementation of lattice-coded PNC. Driven by this issue, recent   interest and effort has been  devoted to
 binary/nonbinary channel-coded PNC.}

\begin{figure*}[!tbp]
\center\vspace{-2mm}
\includegraphics[width=6.1in,height=3.36in]{{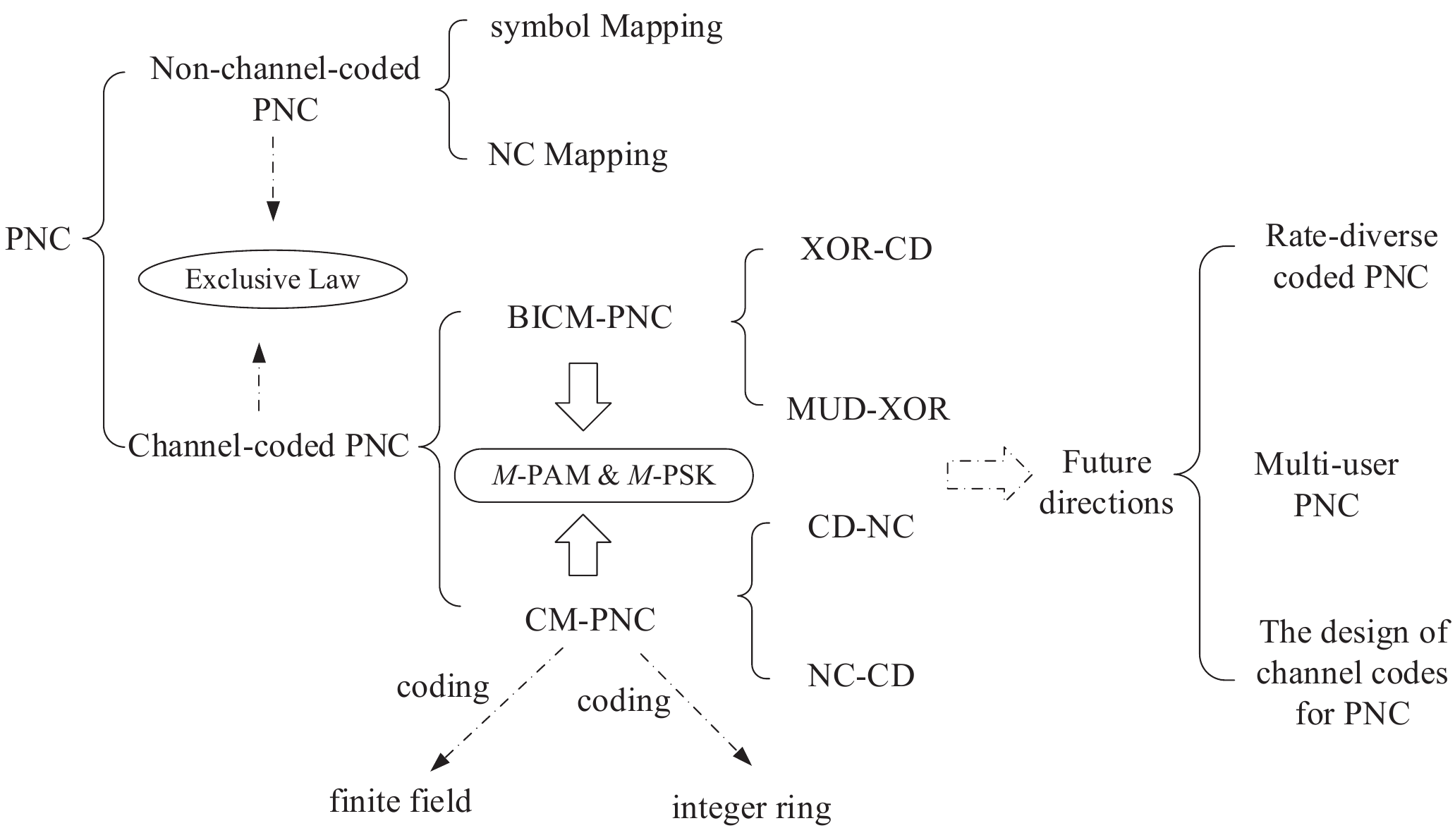}}
\vspace{-0.2cm}
\caption{Organization of the paper.}
\label{fig:Fig.4}  
\end{figure*}
%
%

To  our knowledge, the research of nonbinary channel-coded modulation
PNC (CM-PNC) falls into three general methods, i.e., network coding-based channel decoding (NC-CD), multiuser
complete decoding (MUD-NC), and channel
decoding for nonbinary physical-layer network coding
(CD-NC) \cite{PING2, NCCD}. Accordingly, the binary channel-coded, i.e., bit-interleaved coded
modulation (BICM), counterparts
of NC-CD, MUD-NC, and CD-NC are XOR-based channel decoding
(XOR-CD), MUD-XOR, and CD-XOR,  respectively \cite{LLR, PING1, ZHANG}, as summarized in Fig. 2.  Note that XOR is a form of NC which is usually used to obtain bitwise XOR messages.
Moreover, an interesting phenomenon is that MUD-NC outperforms NC-CD and CD-NC in the
low SNR regime in terms of the achievable rate, and vice versa in the high SNR regime. This comparison also holds for their binary channel-coded counterparts.

NC-CD/XOR-CD first maps superimposed channel-coded
symbols from two users to NC symbols before explicit channel decoding.  Thus, the relay can directly decode channel-coded NC message  without knowing individual
user messages. By contrast,
MUD-NC/MUD-XOR  first decodes two user messages given the superimposed
signals by MUD, and then performs NC mapping on two decoded messages \cite{LIEW}. Unlike the previous schemes, CD-NC/CD-XOR first maps superimposed signals to a transmit symbol pair and then performs channel decoding.  Afterwards, we can obtain NC message based on the decoded transmit pairs.


 For XOR-CD,  \cite{non} developed  XOR-CD
with non-uniform PAM constellation to facilitate the use of
BICM for PNC. While for PSK modulation, \cite{non2} investigated the design the optimal bit
mapping of PSK symbols for BICM-PNC. Moreover, \cite{PING1} showed that the achievable rate of XOR-CD  is smaller than that of its
NC-CD counterparts even under Gray-mapped high-order modulations, and proposed an  iterative XOR-CD to narrow this rate gap.
For NC-CD,  \cite{NCCD, PING1} considered low-density-parity-check (LDPC) coding aided PNC over the finite fields $\mathbb{F}_3$, $\mathbb{F}_{2^2}$ and $\mathbb{F}_5$. Nevertheless,  different superimposed signals before decoding are mapped to the same NC symbol degrades  the performance of channel decoding. For this reason, \cite{PING2} proposed CD-NC   to overcome this issue. CD-NC employs nonbinary LDPC codes over  the integer ring   $\mathbb{Z}_M$ rather than the finite field, and    a novel
generalized   decoding algorithm  was developed accordingly.

In this article, we give a general overview of channel-coded PNC in TWRC. We first present the relevant system models and the NC mapping for PNC. Afterwards, we describe the main encoding and decoding schemes for channel-coded PNC. With simulation results, we illustrate the  decoding performance of different channel-coded PNC schemes over AWGN and Rayleigh fading channels.
To the best of our knowledge, this is the first tutorial touching upon \textit{channel-coded PNC} for the relay networks.
Interested readers are referred to the references for more comprehensive treatments of this promising topic.

\section{System Model}\label{sect:system-model}

In the multiple-access phase of a TWRC, two users encode their source messages using the same channel code to yield two codewords, respectively. Then two signal vectors are created    by $M$-ary modulation on the two codewords, respectively. The two modulated vectors are then simultaneously transmitted  to relay R.
It is pointed out that the goal at the relay is to decode the NC message from the superimposed signals. The NC message is  then encoded for broadcast to the two users.
 The key point in PNC is how to decode the NC message from the superimposed signals in the first phase.

%
In such a channel-coded PNC, the desired NC vector is a valid codeword of the encoder and can be decoded with the same decoder. We consider $M$-PSK and $M$-PAM modulations to study at user nodes. The cardinality of both constellation sets of the two users are the same. As a result,  we have the total $M^2$ joint symol pairs for transmission by treating two user symbols as a transmission pair.  Accordingly,  the receiver can form a   superimposed constellation set, in which each element is a linearly superposition of two modulated symbols weighted by their user-to-relay channel coefficients, respectively.

In general, in a TWRC network over fading channels,  each transmission pair generates a superimposed signal weighted by fading coefficients. Thus, the superimposed set has $M^2$ elements and each element  is distinct.
    On the other hand, in  a TWRC  over AWGN channels where  two user-to-relay channel coefficients are the same, the number of elements in the superimposed set  is less than $M^2$. This is because that multiple transmit pairs  may lead to the same superimposed signal.
    Recall that the goal of PNC  is to compute the NC message out of the superimposed signal at the relay.  It turns out that we should give the NC mapping such that a unique NC symbol can be derived from the superimposed signal  without ambiguity.

\subsection{Non-channel-coded PNC}

In non-channel-coded PNC, we treat an NC mapping as feasible as long as the mapping satisfies the \emph{exclusive law} \cite{non,non2}.  Each of the two users can be ensured to retrieve the message of the other user given its self-information and the received NC message in  the broadcast phase. For example, for $M$-PAM based  PNC, a feasible NC mapper
is the modulo-$M$ arithmetic of two transmit symbols over integer ring $\mathbb{Z}_M$. With respect to $M$-PSK based PNC, a feasible mapper is an addition of two transmit symbols  over the finite field $\mathbb{F}_{2^r}$,  $2^r=M$. Note that  both the demappers can be performed with different $M$-ary decoding coefficients  which belong to a set of non-zero divisors. In this way, we can obtain a unique NC symbol derived from the superimposed symbol which is generated even by multiple transmission pairs.
The choice of decoding coefficients has been intensively  studied in \cite{LONG1}.

It is pointed out that for some channel-coded PNC, it is not strictly required for the NC mapping that all mapped NC symbols cannot have   mapping ambiguity. This is because that channel decoding can recover the corrupted NC symbol of ambiguity in some extent. Even so, the ambiguity in
the channel decoding still undergoes a rate loss theoretically. For this reason, the channel-coded PNC
abides by the same design rule of the NC mapping as non-channel-coded PNC.
We next delve into the details of BICM-PNC and CM-PNC.

\begin{figure}[t]
\centerline{
\includegraphics[height=0.9\columnwidth]{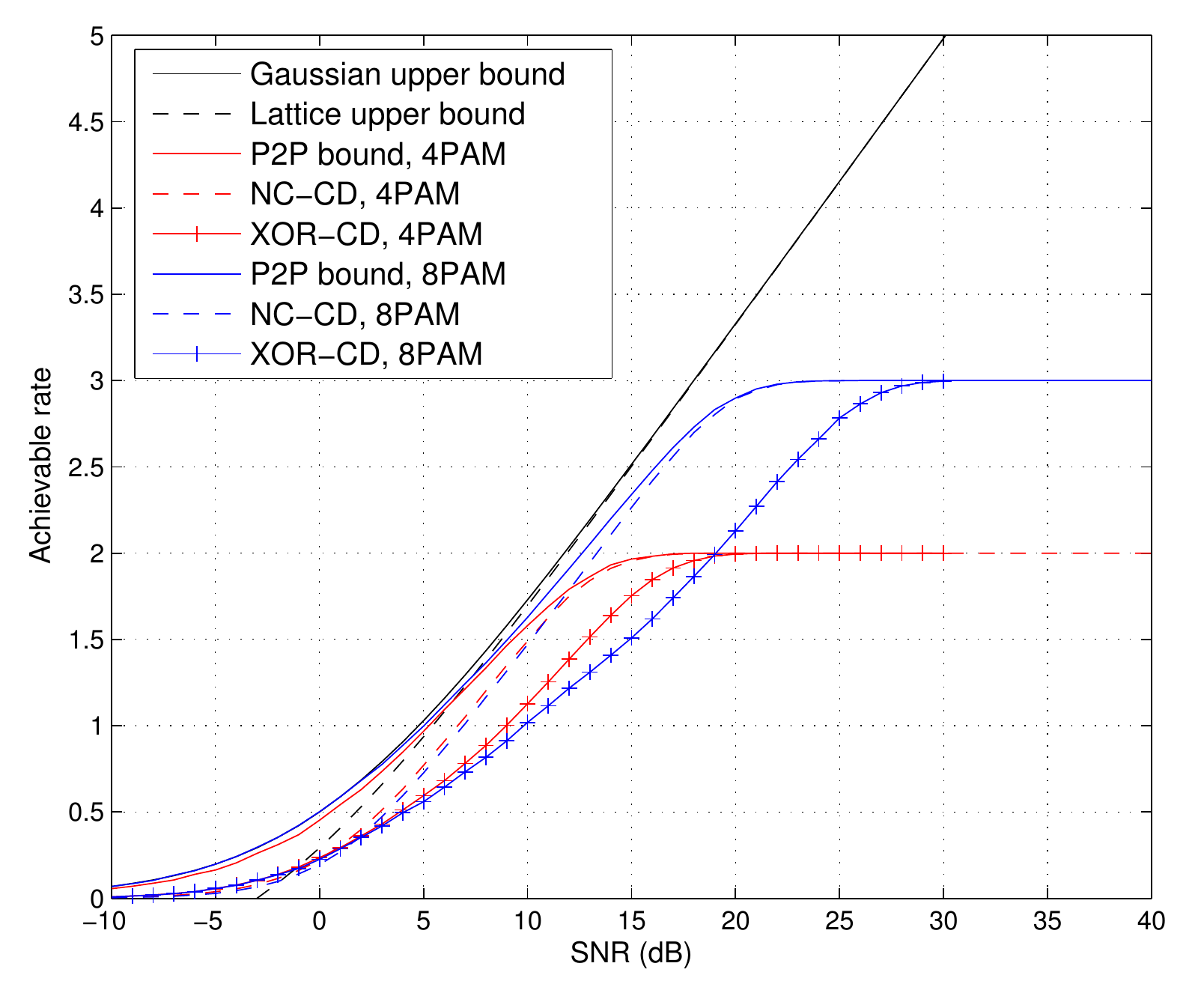}}
  \caption{PNC-rate comparison of XOR-CD and NC-CD over AWGN channels.}
  \label{fig:RATE}
\end{figure}
\subsection{BICM-PNC}
In practice, the decoding complexity of CM-PNC using nonbinary codes  is prohibitively  increased   as  the size of coding symbol increases.
Thus, to achieve high spectral
efficiency and low complexity, we consider BICM-PNC with high-order modulation as an alternative to lattice-coded/nonbinary-coded CM-PNC. For BICM-PNC, we have to select the proper mapping at the user nodes and the NC mapping at the relay. In some channels (e.g., AWGN channels),
 one received symbol in the superimposed constellation  may be associated with more than
one transmit symbol pair, which may lead to different NC symbols, i.e., mapping ambiguity. From this view of point, we should carefully determine
  bit mappings of high modulations to avoid ambiguity.

Consider XOR-CD and MUD-XOR for BICM-PNC. For  $M$-PAM based XOR-CD,
a non-uniform PAM  with modified spacing between
PAM constellation points was proposed as legitimate bit mappings \cite{non}.
For $M$-PSK modulations, \cite{non2} carefully analyzed the symbol mapping
at user nodes,  and proposed a semi-Gray mapping to achieve the best performance in the high SNR regime.

Moreover, BICM capacity and CM capacity for
Gray-mapped point-to-point communication system are almost
the same. By contrast, for both $M$-PSK and $M$-PAM PNC systems, we found that XOR-CD and MUD-XOR suffer an appreciable capacity loss from their CM counterparts, i.e.,
NC-CD, and MUD-NC,  respectively, even with Gray-mapping modulations \cite{PING1,PING2}.    To narrow this loss, we proposed an iterative decoder for BICM-PNC where the messages between bits-to-symbol PNC mapper and channel decoder are further iteratively exchanged and updated \cite{PING1}.

\subsection{CM-PNC}
Although XOR-CD is a simple channel-coded PNC scheme, it has a significant rate gap from nonbinary-coded
CM-PNC with high  modulations, as shown in Fig. 3. We consider $M$-PSK based and $M$-PAM based CM-PNC with the corresponding $M$-ary channel coding schemes, respectively.
Similarly, the NC mapping in CM-PNC needs to satisfy that the NC vector of two user codewords is still a valid codeword.  CM-PNC is also applicable for a TWRC, in which two users adopt $M^2$-QAM modulations, since any QAM signal can be treated as
  two $M$-PAM orthogonal real and imaginary parts. We can directly apply the mapping and coding methods in each part.

In CM-PNC, we emphasize that the operation of channel coding should be consistent with the NC mapper. Recall that the NC mapper  for $M$-PSK based  PNC is an addition of two user symbols over the finite field  $\mathbb{F}_{2^r}$, and for $M$-PAM based  PNC is a mod $M$ operation of two user symbols over the integer ring $\mathbb{Z}_M$ \cite{LLR,non2}. Based on this observation,
for $M$-PSK based CM-PNC, we adopt $M$-ary channel coding over  $\mathbb{F}_{2^r}$.  On the other hand, for  $M$-PAM based CM-PNC, we adopt $M$-ary channel coding over  $\mathbb{Z}_M$ instead of $\mathbb{F}_{2^r}$. In this way, we can keep the  consistency between the decoding operation and the NC mapping operation, which involve multiplications and additions. Also,  decoding coefficients used in the NC mapping   are selected from non-zero divisors in $\mathbb{F}_{2^r}$ and $\mathbb{Z}_M$, for $M$-PSK and $M$-PAM based CM-PNC,  respectively.  Such consistency  is essential for the NC decoding to extract the desired messages at two users.
{We summarize the NC mapping rules for channel-coded PNC systems as follows}:
\begin{itemize}
\item[1)] NC mapping must satisfy the \emph{exclusive law}.
\item[2)] NC mapping cannot incur ambiguity, i.e.,  each superimposed symbol is mapped into a NC symbol uniquely.
\item[3)] The entire mapped NC codeword is a valid codeword of the channel code adopted at the user nodes.
\end{itemize}

\section{Decoders of Channel-coded PNC}

%

\subsection{XOR-CD for BICM-PNC}

Recall that we have two decoders, i.e., MUD-XOR and XOR-CD for BICM-PNC.
  In specific,   XOR-CD  first performs  symbol-by-symbol bitwise XOR to obtain NC symbol from the received signals since  binary channel code is used at two user nodes. Then, the symbol-to-bits demapper computes the soft information about the XOR bits from the NC symbol for the channel decoder.  Note that if multiple possible transmit symbol pairs results in the same received signal,  it may generate different bitwise NC symbols. In such a case, an ambiguity occurs and useful information about the NC symbol is lost, in that we cannot distinguish  which of the NC symbols is correct. As a result, we have to carefully design the transmit constellations of two users to avoid ambiguity \cite{PING1,PING2}.

For example,   Fig. 4(a) shows two different Gray-mapped 8PSK constellation sets $\chi_A$ and $\chi_B$ for two users respectively, where  $\chi_B$
has a  rotation of $\pi/8$ from $\chi_A$. In this case, we have  64 elements in the superimposed set at the relay. { Each element is generated by a unique transmission pair and a unique NC symbol can be determined from the element. It implies that  the constellation sets are applicable for XOR-CD. Moreover, consider an XOR-CD where  the two users employ the same set $\chi_A$ or $\chi_B$ in  Fig. 4.  The superimposed set has 33  elements because some transmission pairs yield the same superimposed element.} Fortunately, from Fig. 4 (b), we see that the ambiguous transmission pairs still yield the same bitwise NC message. Therefore, we conclude that this strategy can also be used for XOR-CD.

  %
%
 %
%

\begin{figure}[t]
    \centering
    \subfigure[Two different constellation sets  for two users respectively, where the dots denote the modulated signals. Two constellations have a angle rotation of $\pi/8$ from each other.]{
    \begin{minipage}{8cm}
    \centering
        \includegraphics[width=3.4in,height=1.6in]{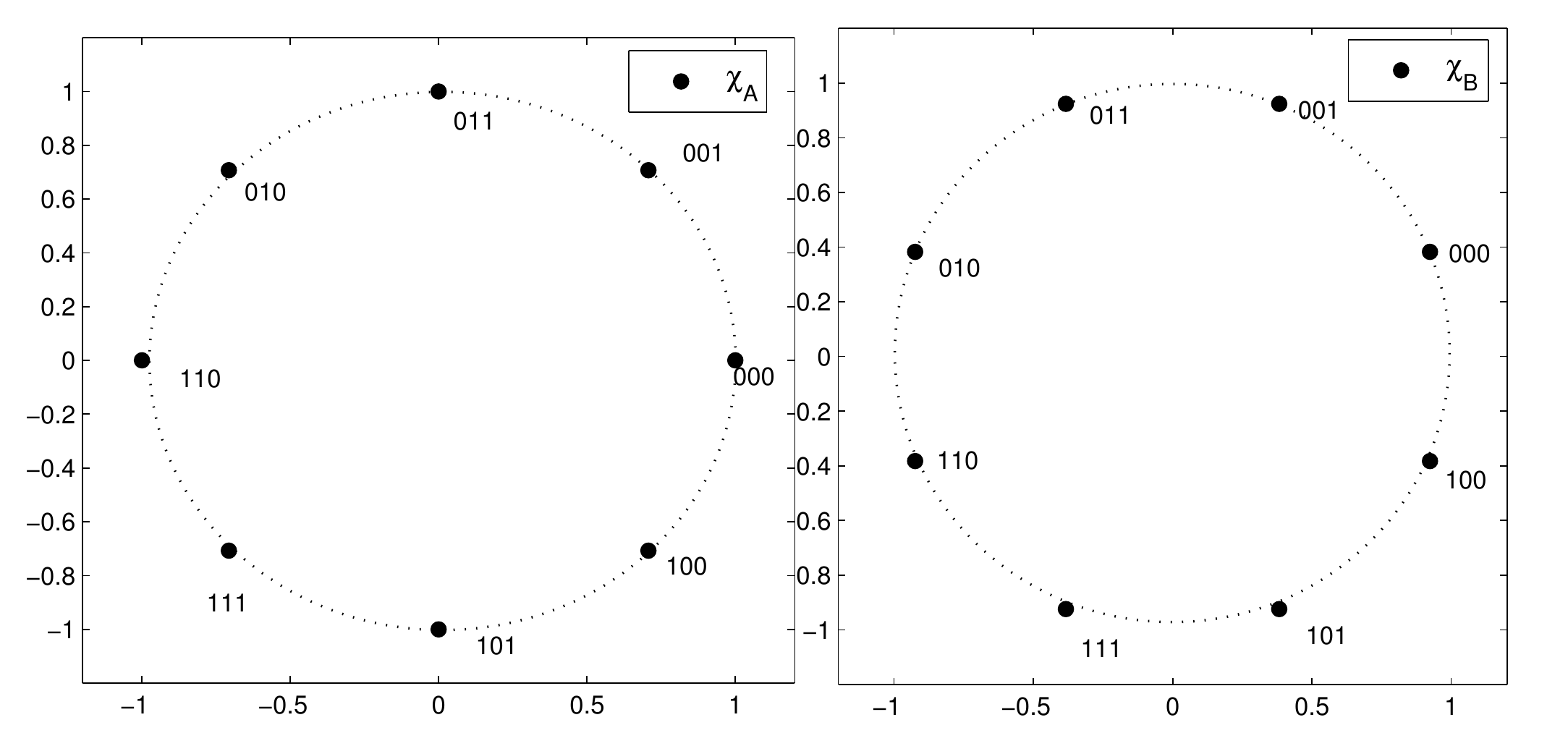}\\
    \end {minipage}
    }

    \subfigure[Received constellation  from the same user constellations]{
    \begin{minipage}{8cm}
    \centering
    \includegraphics[width=3.2in,height=3.0in]{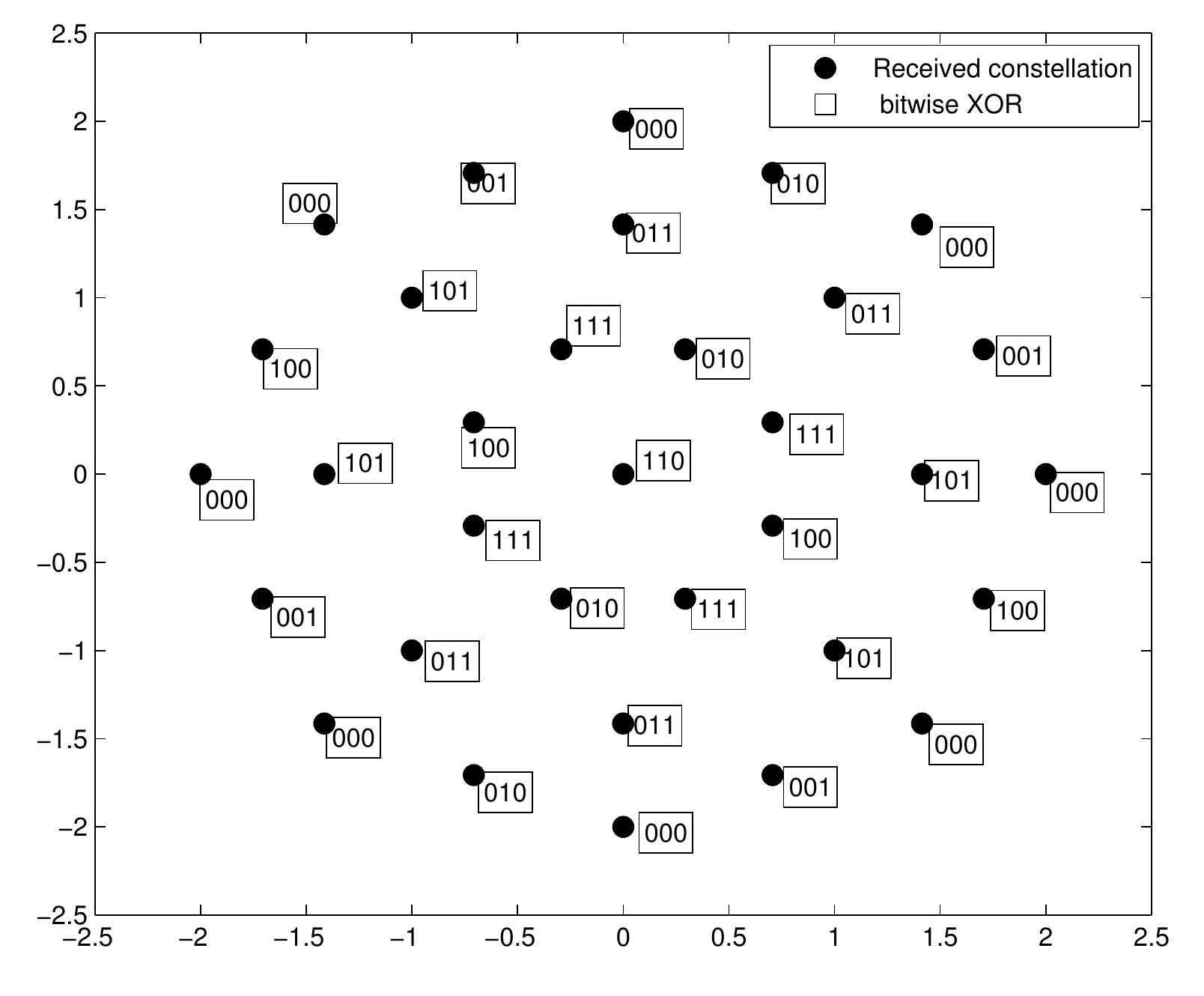}\\
    \end {minipage}}
    \caption{Gray-mapped 8PSK constellation sets and the bitwise NC symbols in channel-coded PNC.}
    \label{fig:fig7}
\end{figure}




\subsection{Iterative XOR-CD for BICM-PNC }
For BICM, different bit-to-symbol mappings of the constellation lead to different channel capacities. It can approach the CM capacity if the bit positions in one symbol are considered as independent. The BICM capacity with Gray mapping is close to the CM capacity for point-to-point systems.  It is not necessary to introduce iterations between channel decoder and symbol-to-bits demapper due to little gain achieved by iterations.
Nevertheless, for XOR-CD with Gray mapping at two users,   the demapped XOR bits within  NC symbols from the superimposed constellation at the relay are not completely Gray-mapped or independent. This can be exemplified from Fig. 4 (b) that the XOR bits within the received constellation  for 8PSK XOR-CD are not Gray-mapped. Based on such an alarming observation,   we consider outer iterations between the binary channel decoder and the NC symbol demapper, which   outperforms its non-iterative counterpart. The performance gain is verified by extrinsic information transfer chart
analysis and distance spectrum analysis \cite{PING1}.  Also,  we indicate that the information capacity of iterative XOR-CD  can approach that of the NC-CD counterpart.

%
\subsection{NC-CD for CM-PNC}
Similar to XOR-CD, NC-CD deals with a symbol-by-symbol NC mapping and then performs channel decoding. As mentioned before, two users  employ the same  nonbinary channel coding  for $M$-PAM   and $M$-PSK  based CM-PNC. In such scenarios,  NC-CD tries to decode multiple NC vectors with different $M$-ary coefficients from the received signal, which enhances the probability for successfully decoding.

As an example, we elaborate the NC-CD for $M$-PAM based CM-PNC. Given decoding coefficients,  we first compute the initial $M$ soft messages for each NC symbol. The computed messages are then fed into a conventional sum-product algorithm (C-SPA) because the entire NC vector is a codeword of the nonbinary coding.   Note that there are total $M^2$  decoding coefficient pairs in NC mapping.
It implies that we can implement C-SPA for one NC
vecotr with a pair of coefficients at most $M^2$ times, leading to the increased decoding complexity, as only one of the successfully decoded NC vectors suffices for our systems. To address this issue, given the fixed channel coefficients, we can decide the best decoding coefficient pair rather than all the pairs for decoding by maximizing the effective minimum distance between NC mapped symbols \cite{LONG1}.
 Nevertheless, the selection of the decoding coefficients also increases  computational complexity and cannot work well when the channel coefficients vary over the codeword transmission.

\subsection{CD-NC for CM-PNC}

Although the achievable rate of NC-CD is larger than that of XOR-CD, it is still far away from the cut-set bound of TWRC theoretically. For example, consider a TWRC with $M$-PAM modulations over Rayleigh fading channels, we remark that there
exists a distinct rate loss of more than 5 dB between NC-CD  and upper bound
 when $M\ge 4$, for both low and high SNR regimes \cite{PING2}.
To  mitigate the loss, we propose CD-NC with a generalized SPA (G-SPA) algorithm to decode the NC vector. In specific, we first virtually construct a encoder where transmit pair symbols are encoded based on the encoding rule at the users and the superposition rule at the relay. Subsequently, we derive check node update and variable node update in G-SPA according to the constructed encoder.
 Unlike NC-CD, CD-NC  first implements channel decoding for the transmission pair symbols and then performs the symbol-by-symbol NC mapping. In G-SPA, we   first compute   $M^2$ probabilities of $M^2$ transmission pairs from the two users, which are then iteratively updated with the derived update rules. Afterward, we can decide the transmission pair of the highest probability  for each NC symbol, which is finally used for NC mapping  to produce the NC symbol.

Revisiting NC-CD, we present the main differences between NC-CD and CD-NC as follows:
\begin{enumerate}
\item NC-CD first calculates the initial soft messages of the mapped NC symbols, and then performs SPA to directly decode the NC vector. In contrast, CD-NC starts with the decoding for all transmission pairs for each NC symbol, and then performs NC mapping after decoding.
\item  NC-CD is implemented at most $M^2$ times, since  multiple NC vectors with $M^2$ decoding coefficients are possibly decoded, while CD-NC is implemented one time. Nevertheless, considering a single decoding process, C-SPA for NC-CD has lower decoding complexity than G-SPA for CD-NC, because the numbers of updated messages per symbol in C-SPA and G-SPA are $M$ and $M^2$, respectively.

\end{enumerate}

For SPA decoding, the operation in the $d$-degree check node dominates the decoding complexity. As mentioned above,  G-SPA iteratively updates  $M^2$ messages for each NC symbol, with the computational complexity being $\mathcal{O}(d\theta^2)$, $\theta=M^2$. This complexity becomes larger as $M$ increases. In point-to-point communications, we prefer to adopt fast-Fourier-transform (FFT) based algorithm of lower decoding complexity. Similarly,  we proposed two-dimensional (2D)-FFT based BP as a simplified alternative of G-SPA \cite{PING2}. The computational complexity of 2D-FFT-BP is reduced to $\mathcal{O}(d\theta\log_2(\theta))$ from $\mathcal{O}(d\theta^2)$ in G-SPA. Moreover, we also apply the  extended min-sum (EMS)  decoding which only involves additions for CD-NC to avoid the multiplications and divisions for message normalization in 2D-FFT. By doing so, the decoding complexity   can be further reduced to $\mathcal{O}(dn_m\log_2 (n_m))$ and $n_m$ is a configure parameter in EMS \cite{PING2}.

\subsection{MUD-NC/MUD-XOR}

For MUD-NC and MUD-XOR, the relay first demaps two user codewords from the received signals individually. Thus,
 if multiple transmission pairs give to the same superimposed signal, an ambiguity  occurs and the achieved rate is reduced  in that we cannot distinguish two user symbols uniquely. Therefore, a constellation
pair for the two users should be designed   such that each superimposed signal corresponds to a unique transmission pair.
Intuitively, two constellation sets cannot be the same. In fact, for $M$-PSK case, a constellation
pair is required to possess an appropriate angle rotation between them, while for $M$-PAM case, the constellation pair has the different spaces between the constellation points.

For example, consider two different Gray-mapped 8PSK constellations for two users, as shown in Fig. 4(a). Then the resulting superimposed set has 64 elements, and each corresponds to a unique transmission pair, which is suitable for MUD-XOR/MUD-NC. {However, if both users adopt the same set,   some elements in the superimposed set may correspond to more than one transmission pairs.  In other words,  the same constellation set at the users introduces ambiguity without yielding uniquely decodable constellation pair. It implies that we cannot apply this strategy for MUD-XOR/MUD-NC.}

In MUD-XOR, we first compute the soft information of two transmit symbols from the two users. Then, the symbol-to-bits demapper computes the soft information of the two coded bits for their respective binary decoders. After channel decoding, the relay achieves a bitwise XOR message on the two decoded codewords. Moreover, we can further introduce  iterative decoding for MUD-XOR where   extrinsic information from one user decoder can be utilized by the other one user decoder. As a result, the decoding performance of MUD-XOR can be improved.


\begin{figure}[!t]
\centerline{
\includegraphics[height=0.9\columnwidth]{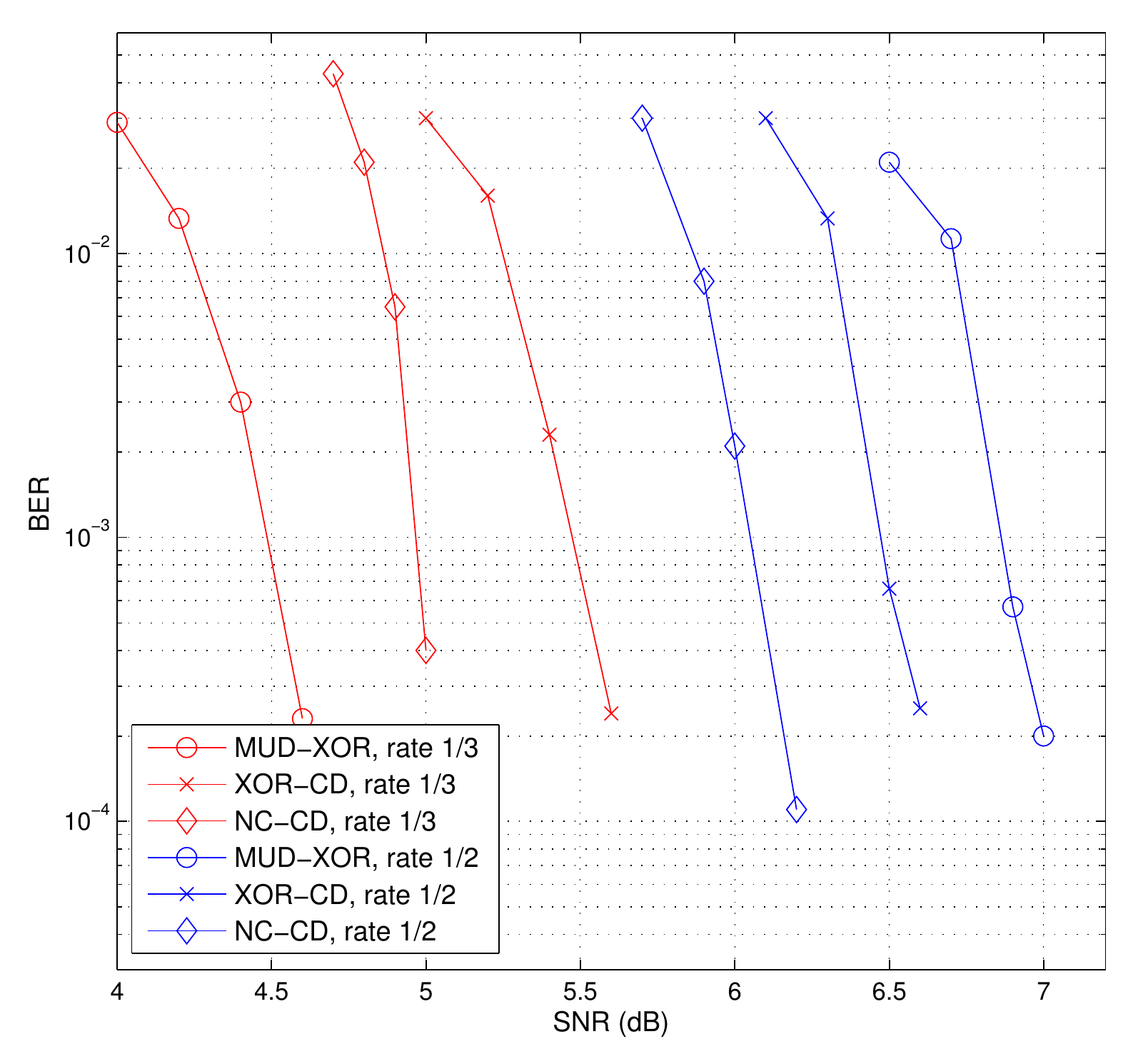}}
  \caption{BER performance of XOR-CD, NC-CD, and MUD-XOR with 8PSK over AWGN channels at rates of $1/3$ and $1/2$.}
  \label{fig:Fig18}\
\end{figure}

\begin{figure}[!t]
\centerline{
\includegraphics[height=0.9\columnwidth]{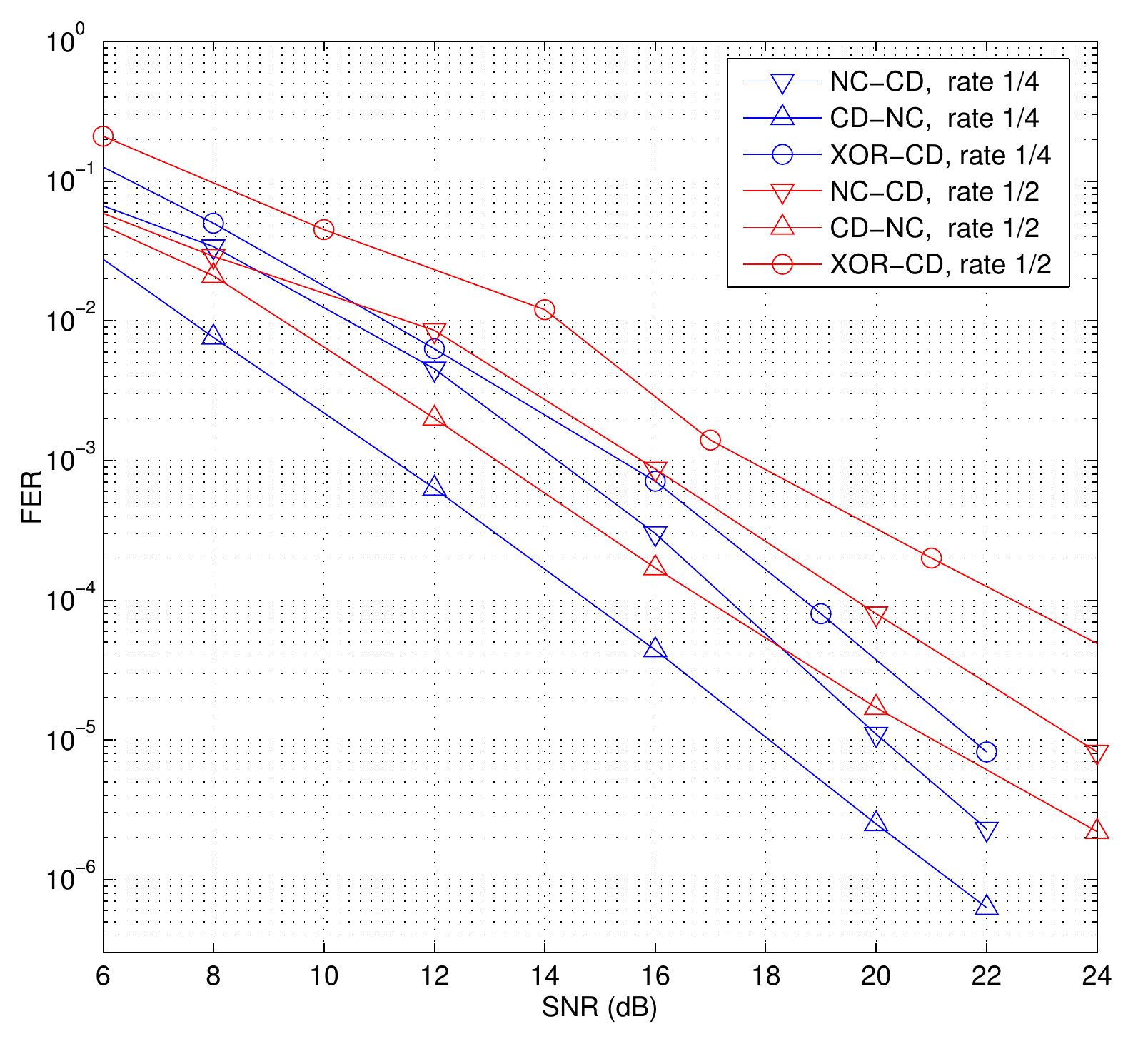}}
   \caption{FER performance of XOR-CD,  NC-CD, and CD-NC with 4PAM over block fading channels at rates of 1/4 and 1/2.}
  \label{fig:BLOCK}
\end{figure}

%
%

\section{Performance Results}\label{sect:simulation}
To evaluate the performance of channel-coded PNC,  we  present  numerical results  of the BICM-PNC and CM-PNC systems under  LDPC  channel coding using VC++ software.
The information length of the codeword is $1032$ for all simulations. The maximum number of decoding iterations is $150$. For a fair comparison, in iterative XOR-CD and MUD-XOR, the numbers of out iterations and channel decoding iterations are assumed as $6$ and $25$, respectively.

In particular, we adopt the binary LDPC codes which were optimized for point-to-point communications, for BICM-PNC, i.e., XOR-CD and MUD-XOR.  For CM-PNC, i.e., NC-CD and CD-NC, we adopt regular nonbinary LDPC  codes. We do not show the performance of CD-XOR since it is a special binary case of CD-NC.

First, we consider PSK-based channel-coded PNC over AWGN channels  where the relay aims for the bitwise XOR of two codewords from the two users.
Fig.~\ref{fig:Fig18} plots the BERs of NC-CD, Gray-mapped MUD-XOR and XOR-CD with 8PSK modulation at code rates of $1/2$ and $1/3$, respectively. We assume a nonbinary LDPC regular code over $\mathbb{F}_{2^3}$ for NC-CD where coding coefficients are randomly selected from $\mathbb{F}_{2^3}$.
We can observe that NC-CD performs the best among the three schemes, and XOR-CD achieves a gain of  $0.8$ dB over MUD-XOR at a rate of $1/2$.
Conversely, MUD-XOR is superior to both XOR-CD and NC-CD  at lower rate of $1/3$.
   The results corroborate theoretical rate analysis that MUD-XOR tends to be better than XOR-CD in the low rate region, while  XOR-CD becomes better in the high rate region. Moreover, NC-CD performs better than XOR-CD for both the code rates due to bits-to-symbol mapping in XOR-CD leads to  information loss as compared to NC-CD.

%

Second, we consider PAM-based   channel-coded PNC over block Rayleigh fading channels.
In strict delay constrained wireless communications,
a codeword with limited length spans a finite number  of $\mathcal{B}$   fading blocks. Our simulated systems set $\mathcal{B}$ to be the typical value  of 4. With Gray-mapped 4PAM modulations, Fig. 6  illustrates the frame error rate (FER) of XOR-CD, NC-CD and CD-NC with the nonbinary channel coding over $\mathbb{Z}_4$  at code rates of 1/2 of and /14, respectively. In this case, NC-CD attempts to decode NC codeword with all possible  coefficients which belong to   non-zero divisors, i.e., $\{(1,3),(1,3),(3,1),(3,3)\}$. It can be observed that CD-NC outperforms NC-CD by more than 2 dB at the FER of $10^{-3}$ for both the code rates. This is because that NC-CD loses the useful information in directly decoding NC message. Also, it is interesting to note that XOR-CD suffers a slight performance loss of 1~dB as compared to NC-CD in this scenario.



\section{Conclusions and Future Research Trends}\label{sect:conclusion-future-line}

\subsection{Conclusions}\label{sect:Conclusions}

In this paper, we conducted a concise overview on the recently proposed channel-coded PNC over TWRC communications and their variant decoders.  To be specific, Sect.~\ref{sect:system-model} introduced the model of channel-coded PNC system and NC mapping design.
After that, we  presented the different PNC decoders for BICM-PNC and CM-PNC in terms of decoding performance and complexity in Section III. 
In Sect.~\ref{sect:simulation}, we showed simulation results to compare the performance of BICM-PNC and CM-PNC. 

\subsection{Future Research Trends}\label{sect:future-line}

\begin{enumerate}[1)]
{
\item
{\bf \textit{Rate-diverse~Channel-coded~PNC:}} Most related works in the literature focus on the PNC where the same channel code is used for two users. A recent study investigated different modulations and the same coding rate for
two users to realize rate-diverse PNC [13]. Nevertheless, little research attention has been paid to the rate-diverse PNC where two users can employ the channel codes of the different coding rates. This study is expected to further enhance achievable PNC rate especially in TWRCs where two user-to-relay channels have different channel conditions.}
 %
{
\item
{\bf \textit{Multi-user~Channel-coded~PNC:}}
Effectively supporting
massive connectivity  is important to ensure that
the forthcoming 5G network can support the Internet of
Things (IoT) functionalities.
Thus, it is significant to investigate  $K$-user PNC communication scenarios, $K\ge 3$.
How to characterize the decoding behavior
for $K$-user PNC  based on the constellation minimum distance [6] is still a challenging task. }

{
\item
 {\bf \textit{Design~of~Channel~Codes for PNC:}} To approach the PNC channel capacity at high SNR regime, prior works have designed  irregular repeat-accumulate codes and  irregular LDPC codes for XOR-CD and NC-CD, respectively. Nevertheless, how to explore variants of the channel codes, e.g., convolutional LDPC codes and spatially-coupled LDPC codes, for CD-XOR and CD-NC is certainly to be expected and deserves further work, especially in low-to-medium SNR regime.}

%

\end{enumerate}

\section*{Acknowledgements}

This work was  supported in part by the NSF of China under Grant 61871132, 61771149, 61671153, the National Funding from the FCT - \textit{Funda\c{c}\~ao para a Ci\^encia e a Tecnologia}, through the UID/EEA/50008/2019 Project, the RNP with resources from MCTIC under Grant 01250.075413/2018-04, the \textit{Centro de Refer\^encia em Radiocomunica\c{c}\~oes} - CRR project of the \textit{Instituto Nacional de Telecomunica\c{c}\~oes} (Inatel), Brazil, the Brazilian National Council for Research and Development (CNPq) under Grant 309335/2017-5, the Natural
Science Foundation of Fujian Province under Grant 2019J01223, the Open Research Fund of National Mobile Communications Research Laboratory, Southeast University (No. 2018D02), the Training Program of FuJian Excellent Talents in University (FETU),  the Guangdong Province Universities and Colleges Pearl River Scholar Funded Scheme under Grant 2017-ZJ022, the Science and Technology Program of Guangzhou (No. 201904010124), and the Research Project of the
Education Department of Guangdong Province (No. 2017KTSCX060). Yi Fang  is the corresponding author.


\begin{thebibliography}{10}
\providecommand{\url}[1]{#1}
\csname url@samestyle\endcsname
\providecommand{\newblock}{\relax}
\providecommand{\bibinfo}[2]{#2}
\providecommand{\BIBentrySTDinterwordspacing}{\spaceskip=0pt\relax}
\providecommand{\BIBentryALTinterwordstretchfactor}{4}
\providecommand{\BIBentryALTinterwordspacing}{\spaceskip=\fontdimen2\font plus
\BIBentryALTinterwordstretchfactor\fontdimen3\font minus
  \fontdimen4\font\relax}
\providecommand{\BIBforeignlanguage}[2]{{%
\expandafter\ifx\csname l@#1\endcsname\relax
\typeout{** WARNING: IEEEtran.bst: No hyphenation pattern has been}%
\typeout{** loaded for the language `#1'. Using the pattern for}%
\typeout{** the default language instead.}%
\else
\language=\csname l@#1\endcsname
\fi
#2}}
\providecommand{\BIBdecl}{\relax}
\BIBdecl

\bibitem{5G}
Z. Ding, X. Lei, G. Karagiannidis, R. Schober, J. Yuan, and V. Bhargava,
``A survey on non-orthogonal multiple access for 5G networks: research
challenges and future trends,'' \emph{IEEE J. Sel. Areas Commun.}, vol. 35, no. 10,
pp. 2181--2195, Oct. 2017,

\bibitem{PNC}
S. Zhang, S. C. Liew, and P. Lam, ``Hot topic: physical layer network coding,'' in \emph{Proc. MobiCom}, Los Angeles, USA, June 2006, pp. 358--365.
\bibitem{NOMA}

T. Yang, L. Yang, Y. J. Guo, and J. Yuan, ``A non-orthogonal multiple-sccess scheme using reliable physical-layer network coding and cascade-computation decoding,'' {\it IEEE Trans. Wireless Commun.}, vol. 16, no. 3, pp. 1633--1645, Mar. 2017.

\bibitem{LIEW}
S. C. Liew, S. Zhang, and L. Lu, ``Physical-layer network coding:
tutorial, survey, and beyond,'' {\it Phys. Commun.}, vol. 6, pp. 4--42, Mar.
2013.

\bibitem{NAZ}
B. Nazer and M. Gastpar, ``Reliable physical layer network coding,'' { \it
Proc. IEEE}, vol. 99, no. 3, pp. 438--460, Mar. 2011.


 \bibitem{LONG1}
L. Shi, and S. C. Liew, ``Complex linear physical-layer network coding,'' {\it IEEE Trans. Inf. Theory}, vol. 63, no. 8, pp. 4949--4981, Aug. 2017.

\bibitem{MIMO}
L. Shi, T. Yang, K. Cai, P. Chen, and T. Guo, ``On MIMO linear physical-layer network coding: full-rate full-diversity design and optimization,'' {\it IEEE Trans. on Wireless Commun.}, vol. 17, no. 5, pp. 3498-3511, Mar. 2018.




\bibitem{LATTICE}
W. Nam, S.-Y. Chung, and Y.  Lee, ``Capacity of the Gaussian two-way relay channel to within $1/2$ bit,'' \emph{IEEE Trans. Inf. Theory},
vol. 56, no. 11, pp. 5488--5494, Nov. 2010.




\bibitem{NCCD}
   Z. F.-Dana and P. Mitran, ``On non-binary constellations for channel coded
physical-layer network coding,'' {\it IEEE Trans. Wireless Commun.}, vol. 12, no. 1, pp. 312--319, Jan. 2013.












\bibitem{PING2}
P. Chen, L. Shi, S. C. Liew, F. Yi, and K. Cai, ``Channel decoding for nonbinary physical-layer network coding in two-way relay systems,'' {\it IEEE Trans. Veh. Technol.},  Vol. 68, no. 1, pp. 628--640, Jan. 2019.


\bibitem{LLR}

R. Y. Chang,  S. Lin, and W. Chung, ``Symbol and bit mapping optimization for
physical-layer network coding with
pulse amplitude modulation,'' {\it IEEE Trans. on Wireless Commun.}, vol. 12, no. 8, pp. 3958-3967, Aug. 2013.



\bibitem{PING1}
P. Chen, S. C. Liew, and L. Shi, ``Bandwidth-efficient coded modulation schemes for physical-layer network coding with high-order modulations,'' {\it IEEE Trans.  on Commun.,}  vol. 65, no. 1, pp. 147-160, Jan. 2017.



\bibitem{ZHANG}

H. Pan, L. Lu, and S. C. Liew, ``Practical power-balanced nonorthogonal
multiple access,'' {\it IEEE J. Sel. Areas Commun.,} vol. 35, no. 10, pp. 2312--2327, Oct. 2017.


%





\bibitem{non}
H. J. Yang, Y. Choi, and J. Chun, ``Modified high-order PAMs for binary
coded physical-layer network coding,''  \emph{IEEE Commun. Lett.},
vol. 14, no. 8, pp. 689--691, Aug. 2010.


\bibitem{non2}
M. Noori and M. Ardakani, ``On symbol mapping for binary physical layer network coding with PSK modulation,''  \emph{IEEE Trans. Wireless Commun.},
vol. 11, no. 1, pp. 21--26, Jan. 2012.




%
%
%
%
%
%
%
%
%
%
%
%
%
%
%
%
%
%
%
%
%
%

\end{thebibliography}

\vspace{5cm}

\section*{Biographies}

\noindent \footnotesize{Pingping Chen [M'15] (ppchen.xm@gmail.com)  received the Ph.D. degree in electronic engineering, Xiamen University, China, in 2013. From May 2012 to September 2012, he was a Research Assistant in electronic and information engineering with The Hong Kong Polytechnic University, Hong Kong. From January 2013 to January 2015, he was a Postdoctoral Fellow at the Institute of Network Coding, Chinese University of Hong Kong, Hong Kong. He is currently a Professor in Fuzhou University, China. His primary research interests include channel coding, joint source and channel coding, network coding, and UWB communications.}

\vspace{0.5cm}

\noindent \footnotesize{Zhaopeng Xie (xzp\_fzu@163.com) received the B.E. degree in electronic
engineering from Hubei University, China, in 2009.
He is currently pursuing the masters degree in electrical engineering with Fuzhou University, China. His
primary research interests include channel coding,
physical network coding, and wireless communications.}

\vspace{0.5cm}

\noindent \footnotesize{Yi Fang [M'15] (fangyi@gdut.edu.cn) received the Ph.D. degree in communication engineering, Xiamen University, China, in 2013. From February 2014 to February 2015, he was a Research Fellow at the School of Electrical and Electronic Engineering, Nanyang Technological University, Singapore. He is currently an Associate Professor at the School of Information Engineering, Guangdong University of Technology, China. His research interests include information and coding theory, spread-spectrum modulation, and cooperative communications. He is the corresponding author of this paper.}

\vspace{0.5cm}

\noindent \footnotesize{Zhifeng Chen [SM'17] (Zhifeng@ieee.org) received his Ph.D. degree in
electrical and computer engineering from the University of Florida in 2010. He is a professor in
the College of Physics and Information Engineering
at Fuzhou University, China. His research interests
include video coding, video transmission, computer
vision, and machine learning.}

\vspace{0.5cm}

\noindent \footnotesize{Shahid Mumtaz [SM'16] (smumtaz@av.it.pt)  received the M.Sc.
degree from the Blekinge Institute of Technology, Sweden,
and the Ph.D. degree from the University of Aveiro, Portugal.
He is currently a senior research engineer with the Instituto de
Telecomunica\c{c}\~oes, Aveiro, where he is involved in EU funded
projects. His research interests include MIMO techniques, multihop
relaying communication, cooperative techniques, cognitive
radios, and game theory.}
%

\vspace{0.5cm}

\noindent \footnotesize{Joel J. P. C. Rodrigues [SM'06] (joeljr@ieee.org) is a professor at the National Institute of Telecommunications (Inatel), Brazil and senior researcher at Instituto de Telecomunica\c{c}\~oes, Portugal. He is the leader of the Internet of Things Research Group (CNPq), Director for Conference Development - IEEE ComSoc Board of Governors, IEEE Distinguished Lecturer, Technical Activities Committee Chair of the IEEE ComSoc Latin America Region Board, the Past-Chair of the IEEE ComSoc TCs on eHealth and on Communications Software, Steering Committee member of the IEEE Life Sciences Technical Community. He is the editor-in-chief of an international journal and editorial board member of several journals. He has authored or coauthored over 750 papers in refereed international journals and conferences, 3 books, 2 patents, and 1 ITU-T recommendation. }

\end{document}